# On the Drag Coefficient Universality of a Rough Grain in Creeping Flow

Si Suo[1,2], Deheng Wei[3,4,†], Budi Zhao[5], Chongpu Zhai[1,*], and Yixiang Gan[4]

[1] State Key Laboratory for Strength and Vibration of Mechanical Structures, School of Aerospace, Xi'an Jiaotong University, Xi'an, China
[2] Department of Civil and Environmental Engineering, Imperial College London, London, UK
[3] Key Laboratory of Ministry of Education on Safe Mining of Deep Mines, School of Resources and Civil Engineering, Northeastern University, Shenyang, China
[4] School of Civil Engineering, The University of Sydney, Sydney, Australia
[5] School of Civil Engineering, University College Dublin, Dublin, Ireland

Controversy exists regarding whether grain morphology reduces or enhances the drag of a single grain in creeping flows; further complication occurs when orientation dependence for aspherical grains comes into play. To quantify influences of shape irregularity and orientation, this study numerically investigates the drag in creeping flows for fractally rough grains depicted by Spherical Harmonics. As the grain surface becomes more angular with increasing relative roughness and fractal dimension, a stronger drag reduction is observed; this observed morphology-dependent reduction indicates Stokes' formula is insufficient. The rotational dependence helps to explain the paradox wherein drag enhancement is always encountered in settling-grain experiments popular in geophysics. For the drag of a given rough grain at various orientations, extracted distributions of the overall drag, and its two components, i.e., skin friction and pressure drag, universally adhere to the Weibull patterns. Moreover, we identify a universal power law between drag coefficients and a newly proposed area-related number, sufficiently taking into consideration both grain roughness by surface area and orientational dependence by projected area perpendicular to the flow direction. This law is primarily controlled by relative roughness, with less sensitivity to fractal dimension. The determined expression enables precise estimation of fluid-particle interactions in upscaling simulations.

† Email address for correspondence: dwei3017@uni.sydney.edu.au.
* Email address for correspondence: zhaichongpu@xjtu.edu.cn.

## 1. Introduction

Understanding the motion mismatch between fluid flow and solid particles is of importance in numerous natural and engineering processes, ranging from the sediment transport in rivers, seepage flow in soils, to the mixing efficiency of industrial procedures. Due to the difficulties in measuring particle shapes with irregularities manifested by multi-scaled roughness, most relevant publications on these fluid-particle interactions consider ideal spheres. Pioneered by Poisson (1831), Green (1833), and Stokes (1845, 1851), the drag coefficient, $C_D$, on spheres has been extensively investigated through experimental, numerical, and analytical approaches for both steady (Oseen, 1910) and unsteady flows (Clift et al., 1978). Given the importance of irregular shapes of natural grains and their orientations, Oberbeck (1876) used Stokes stream function under the assumption of negligible advection to derive the first approximation for ellipsoids at various orientations; Oseen (1915) extended this work by providing higher order approximation to obtain analytical solutions for disks under the flow perpendicular to their flat surfaces. The two classical methods proposed by Stokes and Oseen have later been combined to derive more robust solutions for $C_D$ of aspherical, yet symmetrical grains (Proudman & Pearson, 1957; Breach, 1960; Batchelor, 1969). Nevertheless, the progress of these analytical derivations and their following improvements are rooted in the two minimal conditions: i) low

Reynolds number, $Re$, and ii) ideal symmetrical grains immersed in the flow. The least controversy arises for the sphere in Stokes or creeping flow ($Re \ll 1$), where inertial effects vanish, and the contributions of the pressure and viscous stresses to the drag are analytically formulated: $C_D = f(Re) = \underbrace{8/Re}_{\text{Pressure}} + \underbrace{16/Re}_{\text{Viscosity}}$.

Building on the abovementioned theoretical foundations, numerical and experimental investigations can be conducted to address the limitations of the two simplifications. First, empirically relating $C_D$ to $Re$ remains challenging, even for spheres within the entire subcritical regime (Goossens, 2019). Second, although the ellipsoidal shapes and the orientational dependence have been introduced (Zastawny et al., 2012; Sanjeevi & Padding, 2017), the assumption holds that flow is parallel to one of the three symmetrical planes crossed by two of the three semiaxes. To examine the influence of grain orientation, Sanjeevi and Padding (2017) performed simulations via the lattice Boltzmann method (LBM). Although they found that for prolate spheroids at finite Re the fluid-induced drag and lift forces depend on the inclination angle in the same way as would be predicted in the Stokes regime (Happel & Brenner, 1983), this is not so accurately true for flat oblate spheroids.

Experimental results are preferable in practical applications due to the involvement of all-natural particle shapes. Wadell (1934a, 1934b) realized the importance of the particle area (including flow-direction projected area, $A_p$, and surface area, $S_a$) in determining both the pressure and the friction drags. The drag coefficient $C_D$ was correlated to particle shape parameters, including circularity, $\mathcal{C} = 2\sqrt{\pi A_p}/C$, and sphericity, $\mathcal{S} = \sqrt[2]{36\pi V_p^3}/S_s$, where $C$ is the perimeter of the projected area; $V_p$ is the enclosed volume; and $S_s$ is the volume of the equivalent sphere. Since then, plentiful studies have been dedicated to $C_D = f\left(Re, \underbrace{\mathcal{P}_1 \dots \mathcal{P}_n}_{\text{Shape indices}}\right)$.
Various formulas can be referred to review papers, such as Loth (2008), Bagheri & Bonadonna (2016), Michaelides & Feng (2023), and references therein. However, Hölzer & Sommerfeld (2008) claim that empirical and semi-empirical drag-coefficient models on irregularly shaped grain could yield significant prediction errors, i.e., more than 1,000%, when used for other experimental data sets. The possible explanations are twofold. First, the shape indices in these equations are single-scaled lacking a full-scaled parameter to quantify the whole grain morphology. Wadell himself classified the grain morphology into three aspects: the general form for grain dimensions, the roundness for local corners, and the fine roughness for small features (Wadell, 1932, 1935). Second, in experiments, terminal settling velocity of a natural grain in a quiescent fluid is typically measured. However, during the settlement, the irregular-shaped grain continually adjusts its orientation to the flow direction. This orientation variation affects both $A_p$ and $C_D$. Therefore, results of settling velocity experiments, which yield only one $C_D$ for the corresponding $A_p$, are inherently limited. Moreover, in the creeping regime, the departure of a grain from a sphere has been generally and intuitively considered to increase $C_D$. In the hydraulically smooth regime of low $Re$, roughness is entirely submerged in the viscous layer, exhibiting little impact on skin friction and the drag coefficient (Kadivar et al., 2021); while in rough flow regime, surface roughness could induce drag reduction (Choi et al., 2008). Interestingly, the concept of 'equivalent sand-grain roughness' is widely implemented to characterise roughness in fully rough wall flows at high $Re$ (Nikuradse, 1933).

In this study, we aim to investigate how the grain roughness impacts drag coefficient of a single grain. We conduct over 1,000 Direct Numerical Simulation (DNS) cases in a creeping flow. Various grain shapes are generated based on Spherical Harmonics (SH), exhibiting multi-scale morphology features effectively captured by two compressed shape parameters (fractal dimension and root mean square roughness). Each grain is rotated to 51 orientations with respect to the flow to discuss the morphology and orientational universality in $C_D$. Corresponding contributions of skin friction and pressure drag to the total drag force are explained by specific morphology features. By gaining a deeper understanding of the drag force on rough grains, this research provides valuable insights into particle-fluid interactions in low-$Re$ flows and has implications for unresolved simulations of fluid-particle systems.

## 2. Methods

### 2.1. *Grain shape generation and description*

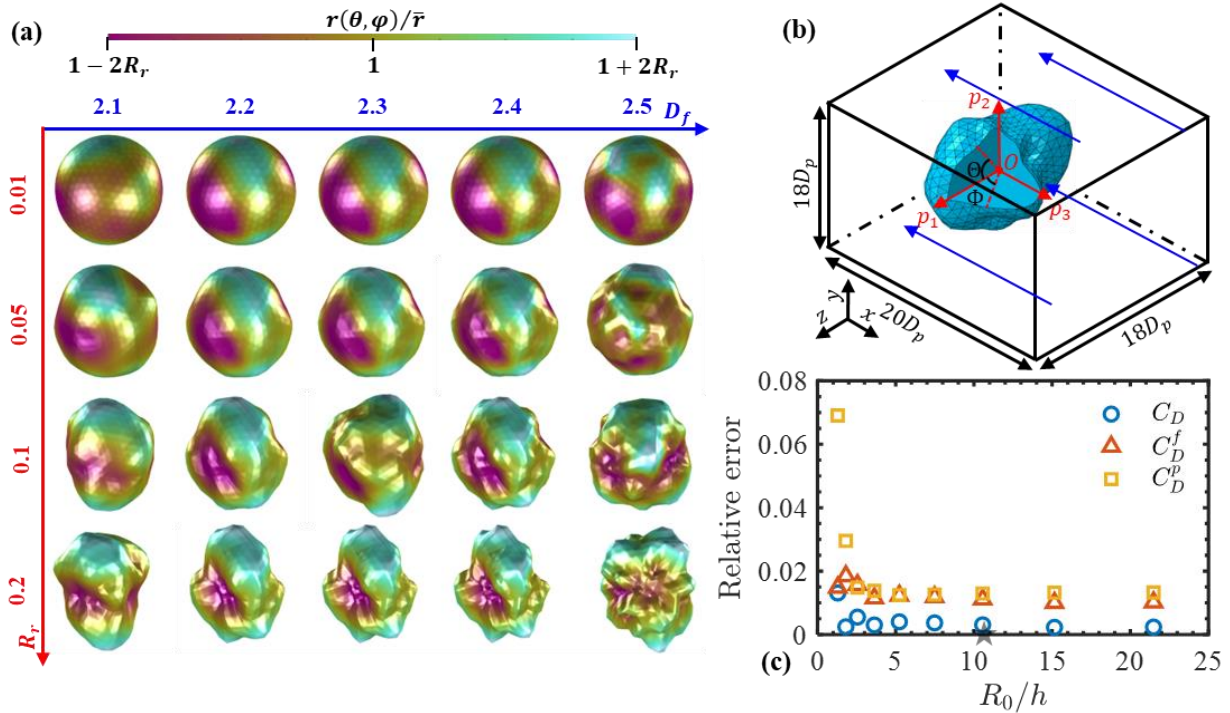

FIGURE 1. (a) Various rough grains in the $R_r$-$D_f$ space. (b) The schematic of a free stream passing a rough grain. (c) Mesh sensitivity study on the benchmark case, i.e., Stokes drag of a sphere. $h$ is the mesh size and $R_0$ is the volume-equivalent sphere radius. The selected mesh size for the systematic study is indicated by the star.

In concurrent studies on drag coefficients of grains, the ideal geometry with fully convex shapes is of high priority. However, realistic grain surfaces usually demonstrate fractality manifested by irregular tortuosity over a wide range of length scales (Barclay and Buckingham, 2009). In this study, we adopt the inverse analysis of Spherical Harmonics (SH) with prescribed power spectrum to generate such multi-scaled morphology features (Wei et al., 2021); the power spectrum is enriched with fractality and root mean square roughness (RMS). A rough grain constructed with star-shaped surficial points can be approximated using the SH function, $Y_n^m(\theta, \varphi)$, via denoting its radial length in spherical coordinate system,

$$r(\theta, \varphi) = \sum_{n=0}^{\infty} \sum_{m=-n}^{n} c_n^m Y_n^m(\theta, \varphi), \quad (2.1)$$

where $\theta \in [0, \pi]$ and $\varphi \in (0, 2\pi]$ are the latitudinal and longitudinal coordinates, respectively, and $c_n^m$ are the SH coefficients of spherical wavenumber $n$ and order $m$. With the help of Parseval's theorem, the power spectrum, $L_n$, at each SH frequency can be calculated by $L_n = \|f_n\| = \sqrt{\sum_{m=-n}^{n} \|c_n^m\|^2}$, with $\|.\|$ being the $L_2$ norm. The average radius length, $\bar{r}$, equalling its $c_0$-

*On the Drag Coefficient Universality of a Rough Grain in Creeping Flow*determined sphere, $R_0$, reads as $\bar{r} \cong R_0 = c_0 Y_0(\theta, \varphi)$, where $Y_0(\theta, \varphi) = 1/(2\sqrt{\pi})$. To quantify how rough grain surface is globally different from $\bar{r}$, the relative roughness is defined as the ratio $R_r = S_q/\bar{r}$, i.e., the RMS, $S_q = \sqrt{(\sum_{n=1}^{\infty}\sum_{m=-n}^{n}\|c_n^m\|^2)/(4\pi)}$, over $\bar{r}$. With $R_r$ and $c_0$ in hand, the grain volume can be correspondingly determined, $V_p = c_0^3(1 + 3R_r^2)/(6\sqrt{\pi})$, which eases the definition of effective grain diameter $D_p = \sqrt[3]{6V_p/\pi} = c_0 \sqrt[3]{1 + 3R_r^2}/\sqrt{\pi}$; the grain surface area is $S_s = c_0^2(1 + \pi R_r^{\pi/2} D_f^{3.874})/20$ (Wei et al., 2021). Meanwhile, the multi-scale features of generated rough grains can be characterized by fractal dimension $D_f$, originated from the logarithmic linear relations between $L_n$ and $n$, i.e., $L_n \propto n^\beta$, where $\beta = -2H$ is the slope of the regression plot of $\log(L_n)$ versus $\log(n)$, and $H$ is the Hurst coefficient used to calculate $D_f = 3 - H$ (Russ, 2013). With increasing $n$, more and more fine surficial details could be depicted. Considering the extremely low $Re = 10^{-5}$, $n_{max} = 15$ with the related cut-off or minimum wavelength, $\lambda(n) = \frac{\pi}{n}\bar{r} = \frac{\pi}{15}\bar{r}$, according to the famous Jean's formula.

As illustrated in Fig. 1 (a), where $c_0 = 2\sqrt{\pi}$ is set to depict a unit sphere with $\bar{r} = 1$, higher $R_r$ and $D_f$ could induce more irregularities in general shapes and local waviness, respectively. For natural granular materials, the value of $D_f$ for quartz sands is mostly in the range $[2.1, 2.3]$, and the value of $R_r$ is around 0.1 (Wei et al., 2021). Extremely fine values of $(R_r, D_f) = (0.01, 2.05)$ could be encountered in artificial spherical particulates, such as glass beads; ultra-high $R_r > 0.15$ and $D_f > 2.4$ could appear in some macromolecules and human brains (Yotter et al., 2010). A series of simulations are conducted for rough grains with $R_r \in [0, 0.2]$ and $D_f \in [2.1, 2.5]$. Notably, we focus here on roughness effect rather than the shape effect (we know they two are actually coupled) and examining the impacts of $D_f$ at different $R_r$ level. Therefore, the total shape of generated particles is close to sphere, whose aspect ratio, elongation, flatness, roundness, sphericity and convexity are demonstrated in Table B1-6 in Appendix B.

In general, as in Fig. 1 (b), drag can be evaluated by a dimensionless coefficient $C_D$ is

$$C_D = \frac{F_D}{\left(\frac{1}{2}\rho_f |\mathbf{u}_\infty|^2 A_p\right)}, \tag{2.2}$$

where $F_D$ is drag force, $\rho_f$ the fluid density, and $|\mathbf{u}_\infty|$ the terminal or free stream velocity of the fluid. To determine $C_D$ for a rough grain, a numerical model is developed in the following section. The analysis of rotational dependence for drag coefficients is performed with two rotations relative to the flow direction: azimuthal rotation $\Phi$ and polar rotation $\Theta$ as in Fig.1 (b). Principal Component Analysis is also utilized to determine major, median, and minor axes with respect to $\mathbf{p}_1$, $\mathbf{p}_2$, and $\mathbf{p}_3$ (Fonseca, 2011), respectively. Notably, minimum and maximum $A_p$ are respectively represented by projected areas vertical to $\mathbf{p}_3$ and $\mathbf{p}_1$. Totally, 49 random rotations plus 2 featured directions ($\mathbf{p}_3$ and $\mathbf{p}_1$) are considered for each morphology. A sensitivity study of the orientation number has been performed, see Appendix A, suggesting that the statistical features of drag can be captured with this orientation number ($N = 51$).

### 2.2. *Governing equations and numerical configuration*

When relative movement between the grain and its surrounding fluid, it experiences a resultant force. The force may not align with the flow due to the asymmetric morphology and can be split into streamwise drag force and spanwise lift force. Here, our study exclusively focuses on the drag force. The total drag force $F_D$ is obtained by a surface-integral of the total stress $\boldsymbol{\sigma}$ on the whole grain domain $\Omega$ projected on streamwise direction $\mathbf{e}_s$,

$$F_D = \left(\oiint_{\partial\Omega} \boldsymbol{\sigma} \cdot d\mathbf{S}\right) \cdot \mathbf{e}_s. \tag{2.3}$$

Here, $\boldsymbol{\sigma} = -p\boldsymbol{I} + \boldsymbol{\tau}$ is a sum of the pressure tensor $p\boldsymbol{I}$ and the shear stress $\boldsymbol{\tau} = [\nabla \mathbf{u} + (\nabla \mathbf{u})^T]$ where $\mathbf{u}$ is flow velocity. Therefore, the total drag force can be considered to be composed of two components, i.e., the skin friction $F_D^f = \left(\oiint_{\partial\Omega} \boldsymbol{\tau} \cdot d\mathbf{S}\right) \cdot \mathbf{e}_s$ and the pressure drag $F_D^p = -\left(\oiint_{\partial\Omega} p d\mathbf{S}\right) \cdot \mathbf{e}_s$. We define $C_D^p$ and $C_D^f$ using $F_D^p$ and $F_D^f$ in the same manner as Eq. (2.2), corresponding to effects of pressure and viscous stresses on drag, respectively.

The twofold integrals in Eq. (2.3) can be obtained by solving the continuity and momentum equations. At $Re = \rho_f U_\infty D_p / \mu_f = 10^{-5} \ll 1$, where $\rho_f$, $\mu_f$, and $U_\infty$ are respectively the density, viscosity, and relative velocity of the fluid, no separation of the boundary layer occurs and the flow field can be adequately described by the Stokes equations. Nondimensionalizing lengths, velocities, and stresses with $D_p$, $U_\infty$, and $\mu_f U_\infty / D_p$, respectively, gives

$$\nabla \cdot \mathbf{u} = 0, \text{ and } \nabla \cdot \boldsymbol{\tau} = \nabla p. \qquad (2.4)$$

Furthermore, forces can be nondimensionalized with $\mu_f U_\infty D_p$. In the following, all variables are in dimensionless forms unless otherwise stated. For the sake of convenience, a scaled drag coefficient is introduced as $C_{\bar{D}} = C_D \cdot Re$, i.e., for smoothed sphere at creeping regime $C_{\bar{D}0}^p = 8$ and $C_{\bar{D}0}^f = 16$. However, for a rough grain, numerical solutions are necessary even at such regime. We examine the scenario of a fully developed free stream passing a single grain. The flow region, as shown in Fig. 1 (b), is a domain of rectangle with $[L_x, L_y, L_z] = [20D_p, 18D_p, 18D_p]$ along, respectively, the streamwise ($x$-axis) and spanwise ($y$- and $z$-axis) directions. This boundary setting allows the stream to flow freely and uniformly near the boundary demonstrating ignorable wall effects, which are verified by the very well agreement between theoretical and simulated $C_D$ of perfectly smoothed spheres in Fig. 1 (c). At the inlet, outlet, and sidewalls, a translational velocity of $U_\infty$ along the streamwise direction is imposed. The grain remains stationary at the domain centre, and a no-slip boundary condition is imposed on the grain surface. In this work, the finite element scheme with the boundary fitted method is adopted. The fluid domain is discretized using more than 600,000 tetrahedron elements with mixed-order shape functions, i.e., Taylor-Hood elements (Taylor & Hood, 1973) to guarantee the divergence stability. The stationary solution of the equation set is obtained by solving a saddle point problem. For the detailed discussion of the weak form and corresponding numerical stabilization techniques, one can refer to Hughes & Mallet (1986) and Hauke & Hughes (1994). The whole numerical scheme is implemented in COMSOL Multiphysics®.

The element size, $h$, is grossly identical to that of generic triangular meshes on grain surfaces. $h = 0.0474 D_p$ is calibrated using the Stokes' drag and sufficient for converged results as confirmed in Fig. 1 (c). The convergency is also tested on the roughest case, with a relative error to the finest mesh less than 3%, see Fig A1 (a), and further validated against a theoretical solution of drag on two ellipsoids, as shown in Fig B1 in Appendix B.

### 2.3. *Impacts of grain morphology on drag*

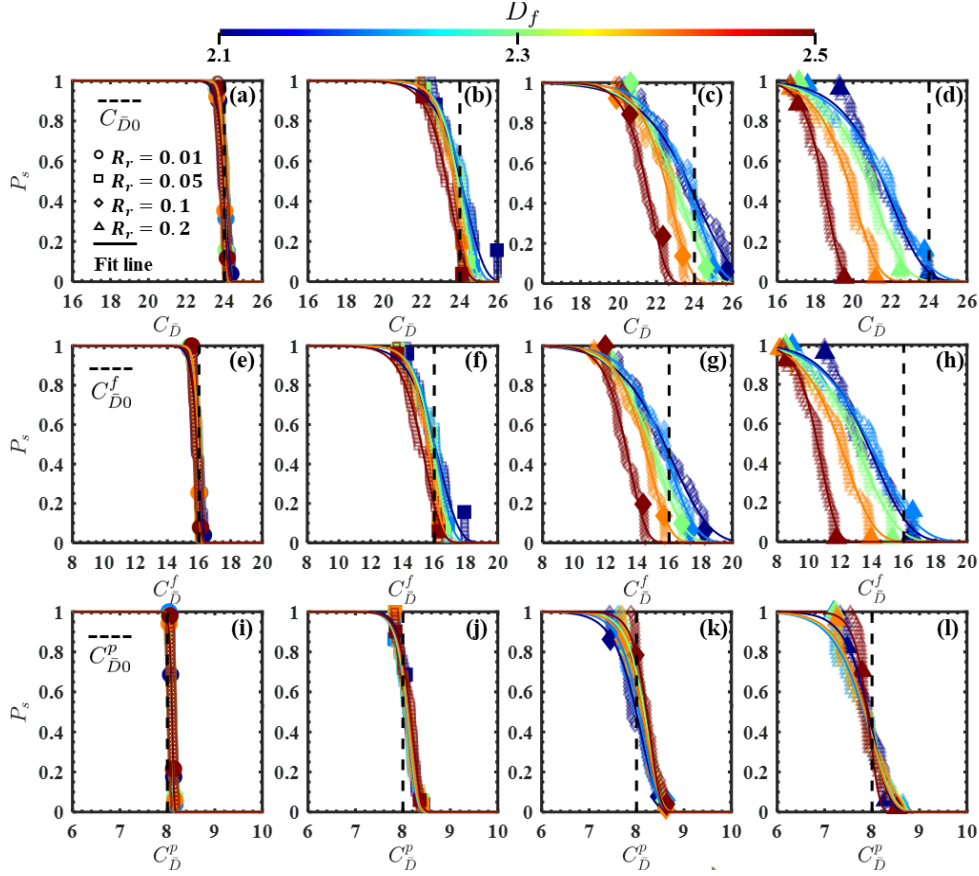

FIGURE 2. Survival probability function $P_s$ of $C_{\bar{D}}$ (a-d), $C_{\bar{D}}^p$ (e-h), and $C_{\bar{D}}^f$ (i-l) at varying rotation angles for each grain morphology in the $R_r$-$D_f$ space. The columns from left to right correspond to $R_r = 0.01, 0.05, 0.1$ and $0.2$. In each sub-figure, the solid ones mark the special rotations with maximum and minimum cross-section areas.

To examine grain morphology effects on the macroscopic drag, we compare in Figs. 2 (a-d) distributions of $C_{\bar{D}}$ collected at varying rotation angles for different grain morphologies, characterized by survival probability function $P_s(C_{\bar{D}}) = P(\tilde{C}_{\bar{D}} > C_{\bar{D}}) = 1 - F(C_{\bar{D}})$ where $F(C_{\bar{D}})$ is the corresponding accumulative distribution function. The drag coefficient is found to be strongly dependent on $R_r$. When $R_r$ is small, i.e., grains are closely spherical, $C_{\bar{D}}$ spans a very narrow range around $C_{\bar{D}0}$, influenced little by $D_f$, suggesting the drag of a small-$R_r$ grain can be approximated using $C_{\bar{D}0}$. When increasing $R_r$, particles become more irregular, and the distribution of $C_{\bar{D}}$ widens and notably shifts towards smaller $C_{\bar{D}}$. This trend persists with higher $D_f$, especially for cases of larger $R_r$. In essence, as the grain becomes more angular with larger $R_r$ and $D_f$, the drag tends to be smaller, resulting in the observed drag reduction effect. Notably, a complete drag reduction, i.e., $C_{\bar{D}} \leq C_{\bar{D}0}$, is observed when $R_r = 0.2$, with corresponding drags overestimated by Stokes' formula.

The grain drag force arises from the combined effects of skin friction, $C_{\bar{D}}^f$, and pressure drag, $C_{\bar{D}}^p$. Figs. 2 (e-h) suggest that $C_{\bar{D}}^f$ follows a similar trend to $C_{\bar{D}}$. However, $C_{\bar{D}}^f$ shows little sensitivity to grain morphology, as illustrated in Figs. 2 (i-l); the mean value of $C_{\bar{D}}^p$ is close to $C_{\bar{D}0}^p$ with a relatively small deviation. Therefore, it can be concluded that the drag reduction is mainly attributed to the decrease in skin friction, whilst the pressure drag exerts a limited influence and even an inverse effect, such as the mean values of $C_D^p$ becomes slightly larger than $C_{D0}^p$ when $R_r < 0.1$. All these macro-observations necessitate a detailed documentation of micro fluid mechanics for potential explanations.

## 2.4. *Grain-scale micromechanics*

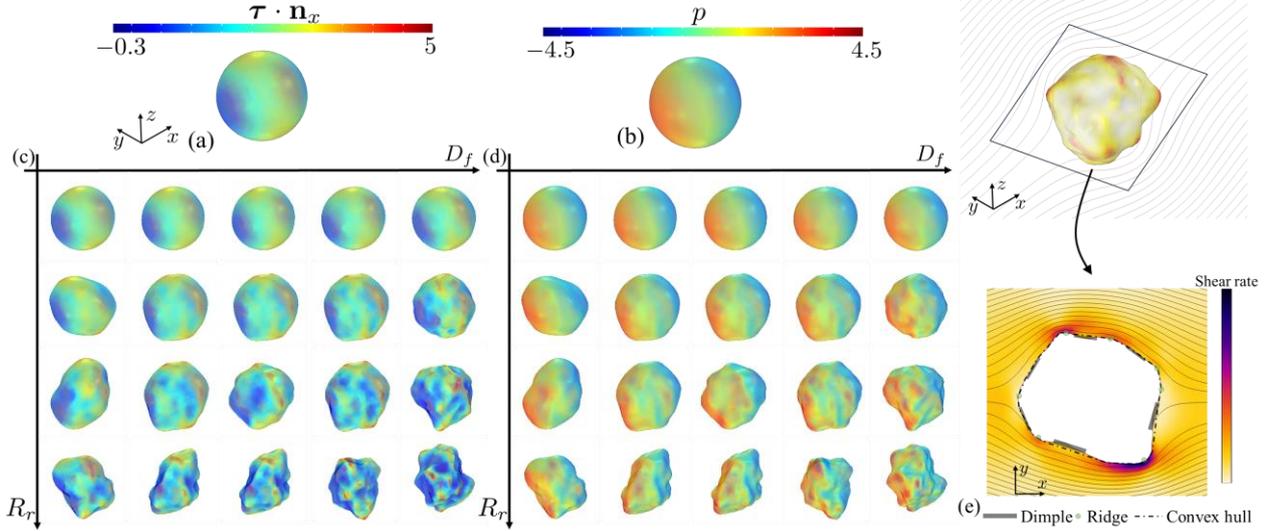

FIGURE 3. The distribution of streamwise shear stress $\boldsymbol{\tau} \cdot \mathbf{n}_x$ (a) and pressure $p$ (b) on a sphere; and $\boldsymbol{\tau} \cdot \mathbf{n}_x$ (c) and $p$ (d) on rough grains. (e) The distribution of streamlines and shear rate for the case with $D_f = 2.4$ and $R_r = 0.1$ and its zoom-in $x$-$y$ plane.

To elucidate the underlying grain-scale mechanisms, we first provide distributions of shear, $\boldsymbol{\tau} \cdot \mathbf{n}_x$, and pressure, $p$, stresses on grain surfaces. For each morphology in Fig. 1 (a), we select the rotation yielding a $C_{\bar{D}}$ closest to the mean value. As in Fig. 3 (a), with a sphere as the reference, $\boldsymbol{\tau} \cdot \mathbf{n}_x$ exhibits a belted distribution with intense values around the equator and relatively weaker values at the front and back polar areas. Such a distribution persists when $R_r$ and $D_f$ are small. On rougher grains, though this global characteristic never vanishes due to the nature of shear flows, a significant shift to the localization is observed, i.e., distributions of $\boldsymbol{\tau} \cdot \mathbf{n}_x$ become more reliant on local geometric features. Specifically, $\boldsymbol{\tau} \cdot \mathbf{n}_x$ is intense at ridges while weak at dimples.

The observed difference in stress distribution between the spherical and rough grains can be explained by Fig. 3 (e). First, ridges and dimples are identified using a convex hull, where vertices indicate the ridges and dimples located between them. For a typical rough grain, Fig. 3(e) demonstrates the streamlines around the grain. Since it is submerged in an open domain, recirculation is not observed in the conducted simulations with low $Re$. However, distinct shear rates are encountered near ridge and dimple regions, as indicated by the zoom-in of Fig. 3 (e), due to the flow deceleration at dimples and acceleration at ridges. Consequently, the skin friction is enhanced at ridges while weakened at dimples. These competing effects collectively determine the overall drag. Notably, the influence areas differ between dimples and ridges. A ridge appears as a wall-mounted singularity and only influences a narrow surrounding area, whereas a dimple covers a wider area where $\boldsymbol{\tau} \cdot \mathbf{n}_x$ is effectively mitigated. Compared with $D_f$ controlling the local features of a grain without changing the general shapes, higher $R_r$ turns out to be more pronounced in leading to a significant drag reduction.

As to $p$ in Fig. 3 (d), though the local extrema emerge due to the roughness, its distribution remains similar to that of a sphere in Fig. 3 (b). Generally, the high pressure can be observed at the upstream surface, while the low pressure occurs at the downstream. The pressure drag is determined by the pressure difference between the upstream and downstream surfaces, which seems to be more dependent on the general grain shape than the local waviness.

## 2.5. *Universal power-law expressions for quantifying drag coefficient*

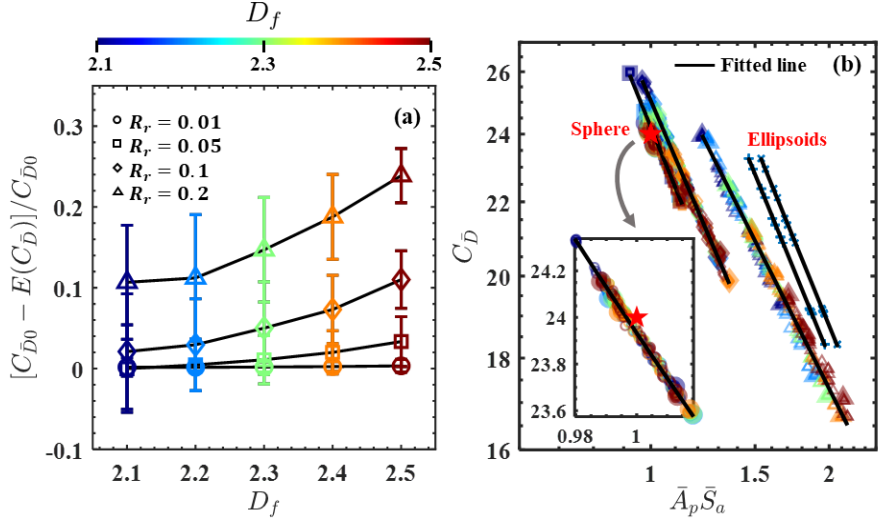

FIGURE 4. (a) Relative difference of $C_{\bar{D}}$ to $C_{\bar{D}0}$ vs. $D_f$ for each $R_r$ in which the error bar is the standard deviation. (b) Log-log plots of $C_{\bar{D}}$ vs. $\bar{A}_p \bar{S}_a$ for each $R_r$, and the insert is a zoom-in plot of $R_r = 0.01$. The solid lines represent the fitting results using a power equation ($y = ax^b$). The meanings of symbols and colours keep consistent with Fig. 2. The red star marks the value of a sphere. Besides, results of two ellipsoids with the aspect ratio of 0.7446 and 0.6258 respectively corresponding to the roughest particle (×) and the particle with smallest aspect ratio (+) are included.

To quantitatively characterize the impacts of grain morphology, a statistical model is developed. Here, we demonstrate that distributions of $C_{\bar{D}}$ for a rough grain at various rotation angles follow the Weibull distribution, of which the survival probability function $P_s(C_{\bar{D}})$ reads

$$P_s(C_{\bar{D}}) = e^{-(C_{\bar{D}}/\lambda)^k}, \qquad (3.1)$$

where $\lambda$ is the scale parameter with $P_s(\lambda) \approx 37\%$, and $k$ the shape parameter which determines the stretch of $P_s(C_{\bar{D}})$. Interestingly, Fig. 2 shows all of $C_{\bar{D}}$, $C_{\bar{D}}^f$, and $C_{\bar{D}}^p$ follow Weibull distributions, regardless of $R_r$ and $D_f$. The coefficient of determination, $R^2$, an index of the goodness-of-fit, exceeds 0.90 for all cases. In general, both higher $D_f$ and $R_r$ could lead to drag reduction, consistent with the finding in rough channel flow. As shown in Fig. 4(a), the drag reduction can reach larger than 20%.

Though the impacts of grain morphology can be universally depicted by Weibullian behaviour, the obtained relationship, i.e., Eq. (3.1) cannot be directly utilized in analytical or numerical frameworks. For instance, both the computed mean value $E$ of $C_{\bar{D}}$ over various orientation angles and the corresponding standard deviation $Sd$ vary largely with $R_r$ and $D_f$, as shown in Fig. 4 (a). Moreover, the rotation-dependence is pronounced for low-$D_f$ grains. An alternative expression with morphology and rotation dependence for the grain drag is desired. As presented above, the drag of a rough grain is determined by its morphology and orientation. Considering that the two factors can be indicated by surface area and projection area, respectively, we propose a compound variable, $\bar{A}_p \bar{S}_a = (A_p/A_s)(S_a/S_s)$, using normalized formulas by those of volume-equivalent spheres. The inverse number of $\bar{S}_a$ is the so-called sphericity that has been frequently used, together with $A_p$, in empirical equations to predict $C_D$ (Ganser, 1993). Surprisingly, a power-law universality exists for all $R_r$ regardless of $D_f$, given as $C_{\bar{D}}(\bar{A}_p, \bar{S}_a) = a(\bar{A}_p \bar{S}_a)^b$. Notably, the universality of this power law can be also proved by the results of two ellipsoids as presented in Fig 4(b). Leith (1987) also proposed a similar equation, $C_{\bar{D}}(\bar{A}_p, \bar{S}_a) = 24\left(\frac{1}{3}\bar{A}_p^{0.5} + \frac{2}{3}\bar{S}_a^{0.5}\right)$, assuming the form and the friction drags in Stokes's law still holds for irregular shapes, which is argued by our DNS simulations in Fig. 3, especially for $\boldsymbol{\tau} \cdot \mathbf{n}_x$.

Noticeably, each solid line of Fig. 4 (b) for a specified $R_r$ includes cases of multiple grains with different $D_f$ and various orientations. All these results collapse onto the same line, demonstrating significantly high goodness of fitting with $R^2$ generally larger than 0.99. The scaling parameter, $a$, is solely controlled by $R_r$; while the exponent, $b$, seems to be constant and less sensitive to $R_r$. This trend again suggests that the Stokes' formula based on the uniformity assumption fails to capture the impact of grain morphology on grain drag. Moreover, local surface features, such as the fine cross-scale roughness play a less important role than that of the relative roughness. This is because at creeping flow regime, the Brinkman screening length—the minimum length—that can be identified by viscous fluid (Koch & Brady, 1985), is longer than the altered length by $D_f$ for utilized $R_r$ in this study.

## 3. Discussions and Conclusions

Looking back to settling-grain-experiment based empirical equations to predict $C_D$ of irregular grains at creeping flows (Michaelides & Feng, 2023), nearly all of them have the same general formulation: $C_D = C_{D0}(1 + \mathcal{F})$ with $\mathcal{F} > 0$ being a function of shape indices. The drag enhancement in experiments is contradictory to the prevalent view—surface roughness equals drag reduction—in both rough channel flow (Choi et al., 2018) and simulations of single rough grains with various rotations. To explain this paradox, we start with the experiment itself of settling grain into static fluid. Although rotated less or more periodically, the gravity-driven grain tends to fall down with its maximum $A_p$ of higher likelihood to be perpendicular to the relative movement direction (Dietrich, 1982). This manifestation could induce lower falling down velocity, $w_f$, of the grain. Considering the experimental manner to determine $C_D = 2(\rho_s - \rho_f) \cdot g \cdot V_p/(\rho_f A_s w_f^2)$, where $\rho_s$ is the density of the grain, and $A_s$ is the projected area of the grain volume-equivalent sphere, $C_D$ is increased for higher roughness. In fact, $A_p$ instead of $A_s$ is much better in such an equation, which has also been applied by many studies, but quantifying the inconsistent values of $A_p$ is nontrivial during the falling down process of the grain with periodic rotations. As demonstrated in Fig. 2 and Fig. 4 (b), by the experimental type of rotation $C_D$ for one specific shape (especially for higher $R_r$) are much lower than the mean value of various rotations. Although the mean value is negatively related to $R_r$, the standard deviation even vanishes at extremely low $R_r$; this facilitates the possibility where $C_D$ of high $R_r$ could be larger than that of low $R_r$, resulting in roughness-induced drag enhancement in experiments. One may argue that in the situation of complete drag reduction, i.e., for both extremely high $R_r$ and $D_f$, $C_D$ is always smaller than $C_{D0}$, and no drag improvement could be encountered. Considering the narrow ranges of natural grains ($R_r < 0.1$ and $D_f < 2.3$), for which settling experiments are conducted and mostly from geoscience community for sedimentary study, complete drag reduction is less likely to happen.

In conclusion, this study comprehensively investigates the impact of grain morphology and rotation on drag at creeping flows. Using Spherical Harmonics, a series of rough grains, with irregularities quantified by relative roughness $R_r$ and fractal dimension $D_f$, are simulated. As the grain surface becomes more angular with increasing $R_r$ and $D_f$, a stronger drag reduction is observed, demonstrating much smaller $C_{\bar{D}}$ than that of a smooth sphere. Aided by rotation dependence, the drag paradox between settling experiments and prevalent view in rough channel flow is successfully explained. This reduction highlights a potential misestimation of rough grain drag when relying solely on Stokes' formula. The micro analysis underscores that the primary cause of drag reduction is decreased skin friction due to flow dynamics around ridges and dimples. Importantly, the drag coefficient distribution for a given rough grain universally adheres to the Weibull pattern. Moreover, a universality, $C_{\bar{D}} = a(\bar{A}_p \bar{S}_a)^b$, between $C_{\bar{D}}$ and a scaled area-related number $\bar{A}_p \bar{S}_a$ involving in the overall grain roughness and orientation, is found. This law is found to be primarily controlled by global shapes, i.e., $R_r$, with minimal sensitivity to finer local waviness, i.e., $D_f$.

The obtained universal expression paves the way for precise estimation of fluid-particle interaction, especially in geophysics and geotechnics. For instance, in the soils, fine particles will migrate with seepage flows, leading to the so-called internal erosion which is one of the most common causes of earthen dam failures. Accurate estimation of grain drags will improve our understanding of the underlying migration mechanisms and help predict the critical states.

**Appendix A. Sensitivity Study of Mesh Size and Orientation Number**

We perform a mesh dependence study for the roughest particle ($D_f = 2.5$ and $R_r = 0.2$). The case with $h = 0.0113 D_p$ is taken as a ground truth. As shown in Fig. A1(a), numerical convergency is guaranteed for $h = 0.0474 D_p$ with the relative error less than 3%. We then perform a sensitivity study of orientation number, and two orientation numbers $N = 51$ and $N = 87$ are investigated. Their statistical results are compared in Fig. A1(b) and the fitting parameters for Eq. (3.1) are presented in Table A1.

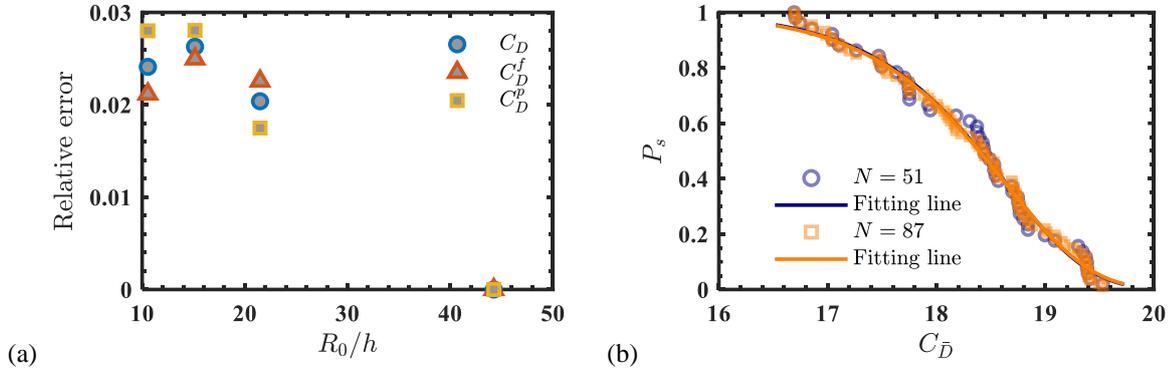

FIGURE A1. Sensitivity study of (a) mesh size and (b) orientation number for the roughest particle ($D_f = 2.5$ and $R_r = 0.2$).

Table A1 Fitting parameters

| $N$ | $\lambda$ | $k$ |
|---|---|---|
| 51 | 25.1889 | 18.6689 |
| 87 | 24.7258 | 18.6673 |

**Appendix B. Equivalent Ellipsoids**

We determine the equivalent ellipsoids for the rough particles shown in Fig. 1 (a) by 3D shape analysis based on the covariance matrix constructed by spatial coordinates of all particle points. The relevant shape parameters including aspect ratio, elongation, flatness, roundness, sphericity, and convexity are calculated accordingly, see Table B.1-6. The quantification method can be referred to Zhao and Wang (2016).

Table B1 Aspect ratio

| $R_r$ \ $D_f$ | 2.1 | 2.2 | 2.3 | 2.4 | 2.5 |
|---|---|---|---|---|---|
| 0.01 | 0.9687 | 0.9644 | 0.9652 | 0.9663 | 0.9796 |
| 0.05 | 0.8561 | 0.8322 | 0.8351 | 0.8411 | 0.8734 |
| 0.1 | 0.7350 | 0.7325 | 0.8164 | 0.7546 | 0.84106 |
| 0.2 | 0.7200 | 0.6258 | 0.64234 | 0.6603 | 0.7446 |

Table B2 Elongation

| $R_r$ \ $D_f$ | 2.1 | 2.2 | 2.3 | 2.4 | 2.5 |
|---|---|---|---|---|---|
| 0.01 | 0.9770 | 0.9815 | 0.9807 | 0.9800 | 0.9873 |
| 0.05 | 0.9045 | 0.9057 | 0.9015 | 0.8977 | 0.8979 |
| 0.1 | 0.8875 | 0.8280 | 0.9516 | 0.8181 | 0.9452 |
| 0.2 | 0.7211 | 0.7426 | 0.7451 | 0.7492 | 0.9382 |

Table B3 Flatness

| $R_r$ \ $D_f$ | 2.1 | 2.2 | 2.3 | 2.4 | 2.5 |
|---|---|---|---|---|---|
| 0.01 | 0.9915 | 0.9826 | 0.9842 | 0.9860 | 0.9922 |
| 0.05 | 0.9465 | 0.9189 | 0.9265 | 0.9370 | 0.9727 |
| 0.1 | 0.8282 | 0.8846 | 0.8580 | 0.9224 | 0.8898 |
| 0.2 | 0.9985 | 0.8427 | 0.8620 | 0.8814 | 0.7937 |

Table B4 Roundness

| $R_r$ \ $D_f$ | 2.1 | 2.2 | 2.3 | 2.4 | 2.5 |
|---|---|---|---|---|---|
| 0.01 | 0.9595 | 0.9524 | 0.9447 | 0.9343 | 0.9210 |
| 0.05 | 0.8410 | 0.8172 | 0.7936 | 0.7670 | 0.7334 |
| 0.1 | 0.7441 | 0.7194 | 0.6813 | 0.6603 | 0.6174 |
| 0.2 | 0.6615 | 0.6327 | 0.6035 | 0.5752 | 0.5295 |

Table B5 Sphericity

| $R_r$ \ $D_f$ | 2.1 | 2.2 | 2.3 | 2.4 | 2.5 |
|---|---|---|---|---|---|
| 0.01 | 0.9986 | 0.9985 | 0.9984 | 0.9982 | 0.9979 |
| 0.05 | 0.9887 | 0.9865 | 0.9833 | 0.9786 | 0.9721 |
| 0.1 | 0.9610 | 0.9534 | 0.9422 | 0.9274 | 0.9070 |
| 0.2 | 0.8814 | 0.8624 | 0.8359 | 0.8013 | 0.7566 |

Table B6 Convexity

| $R_r$ \ $D_f$ | 2.1 | 2.2 | 2.3 | 2.4 | 2.5 |
|---|---|---|---|---|---|
| 0.01 | 1.0000 | 1.0000 | 1.0000 | 1.0000 | 1.0000 |
| 0.05 | 0.9991 | 0.9977 | 0.9946 | 0.9889 | 0.9789 |
| 0.1 | 0.9837 | 0.9748 | 0.9567 | 0.9399 | 0.9110 |
| 0.2 | 0.8988 | 0.8874 | 0.8513 | 0.8068 | 0.7536 |

To further validate our numerical scheme, two ellipsoids corresponding to the particle with the smallest aspect ratio ($D_f = 2.2$ and $R_r = 0.2$) and the roughest particle ($D_f = 2.5$ and $R_r = 0.2$) are selected and named "Ep1" and "Ep2" respectively. Fig. A1 shows the comparison between simulation results and theoretical solutions reported in Oberbeck (1876). The relative error is less than 0.51%.

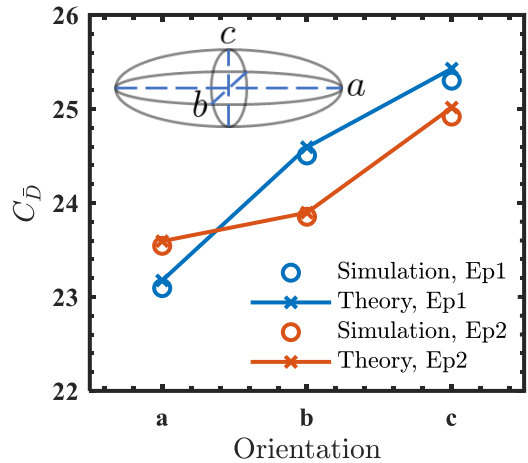

FIGURE B1. Comparison between the simulation and the theoretical solution on drag coefficients of ellipsoids (Ep1 and Ep2) under three principal orientations.

**Acknowledgements.**

This work is supported by the National Natural Science Foundation of China (NSFC, Grant No. 12202342 and the Excellent Young Scientists Fund (overseas) for Dr. Deheng Wei) and by the Australian Research Council (ARC, the Discovery Early Career Award (DECRA), Grant No. DE240101106).

**Declaration of interests.** The authors report no conflict of interest.